**Title:**

Cluster analysis of presolar silicon carbide grains: evaluation of their classification and astrophysical implications

**Short Title:**

Cluster analysis of presolar silicon carbide grains

**Authors:**

Asmaa Boujibar[1], Samantha Howell[1,2], Shuang Zhang[1], Grethe Hystad[3], Anirudh Prabhu[4], Nan Liu[5], Thomas Stephan[6,7], Shweta Narkar[4], Ahmed Eleish[4], Shaunna M. Morrison[1], Robert M. Hazen[1], Larry R. Nittler[1]

**Affiliations:**

[1]Earth and Planets Laboratory, Carnegie Institution for Science, [2]Washington College, [3]Purdue University Northwest, [4]Tetherless World Constellation, Rensselaer Polytechnic Institute, [5]Washington University in St. Louis, Sciences, [6]The University of Chicago, [7]Chicago Center for Cosmochemistry

Corresponding author: aboujibar@carnegiescience.edu

**Abstract**

Cluster analysis of presolar silicon carbide grains based on literature data for $^{12}C/^{13}C$, $^{14}N/^{15}N$, $\delta^{30}Si/^{28}Si$, and $\delta^{29}Si/^{28}Si$ including or not inferred initial $^{26}Al/^{27}Al$ data, reveals nine clusters agreeing with previously defined grain types but also highlighting new divisions. Mainstream grains reside in three clusters probably representing different parent star metallicities. One of these clusters has a compact core, with a narrow range of composition, pointing to an enhanced production of SiC grains in asymptotic giant branch (AGB) stars with a narrow range of masses and metallicities. The addition of $^{26}Al/^{27}Al$ data highlights a cluster of mainstream grains, enriched in $^{15}N$ and $^{26}Al$, which cannot be explained by current AGB models. We defined two AB grain clusters, one with $^{15}N$ and $^{26}Al$ excesses, and the other with $^{14}N$ and smaller $^{26}Al$ excesses, in agreement with recent studies. Their definition does not use the solar N isotopic ratio as a divider, and the contour of the $^{26}Al$-rich AB cluster identified in this study is in better agreement with core-collapse supernova models. We also found a cluster with a mixture of putative nova and AB grains, which may have formed in supernova or nova environments. X grains make up two clusters, having either strongly correlated Si isotopic ratios or deviating from the 2/3 slope line in the Si 3-isotope plot. Finally, most Y and Z grains are jointly clustered, suggesting that the previous use of $^{12}C/^{13}C= 100$ as a divider for Y grains was arbitrary. Our results show that cluster analysis is a powerful tool to interpret the data in light of stellar evolution and nucleosynthesis modelling and highlight the need of more multi-element isotopic data for better classification.

## 1. Introduction

Presolar grains are condensates formed in stellar outflows or explosions during the advanced stages of stellar evolution and now found preserved in meteoritic materials. They can be recognized through their highly anomalous isotopic compositions that differ strongly from material formed in the Solar System (see Nittler & Ciesla 2016; Zinner 2014). These anomalous isotopic compositions for a number of elements, e.g., C and O, are indicators of their formation in stars and provide a variety of information on their stellar sources, from the original Galactic material from which their parent stars formed, to the nucleosynthesis occurring in stars and the mixing of different layers within stars and exploding novae and supernovae. Among the different mineral phases of presolar grains, silicon carbides (SiC) are the most studied, since they can be isolated from meteorites relatively easily by acid dissolution. Their sizes typically range from a few hundred nm up to several micrometers, allowing detailed isotopic characterization by laboratory methods such as secondary ion mass spectrometry.

Previous studies used the isotopic compositions of presolar SiC grains to classify them into different types, potentially corresponding to different stellar formation environments. The majority of presolar SiC grains formed in asymptotic giant branch (AGB) stars, a late stage in the evolution of low- and intermediate-mass (1−8 M$_\odot$) stars. AGB stars are believed to be the parents of ~95% of SiC grains, those classified as mainstream (hereafter MS, representing ~90% of all grains), Y (a few %), and Z (a few %) types (Zinner 2014). The MS grain population was defined as having a C isotopic composition ($^{12}$C/$^{13}$C) between 12 and 100, and $\delta^{29}$Si/$^{28}$Si and $\delta^{30}$Si/$^{28}$Si correlating with a slope of 1.37 (e.g., (Lugaro et al. 1999; Stephan et al. 2020; Zinner 2014) (Fig. 1a). Y and Z grains have Si isotopic compositions deviating from the linear correlation observed for MS grains toward larger $^{30}$Si enrichments relative to $^{29}$Si, with a larger deviation for Z grains than for Y grains (e.g., (Zinner et al. 2006). These compositions were interpreted as the result of formation in low- to intermediate-mass AGB stars of lower-than-solar metallicities, which have enhanced maximum stellar temperatures to allow more efficient operation of the reaction $^{22}$Ne($\alpha$,$n$)$^{25}$Mg and, in turn, higher production of the neutron-rich isotope $^{30}$Si by neutron capture (Hoppe et al. 1997; Lugaro et al. 1999). In addition, Y grains are defined to have $^{12}$C/$^{13}$C > 100, while Z grains have $^{12}$C/$^{13}$C < 100, similar to MS grains. This difference in C isotopic composition was interpreted as the signature of possible cool-bottom processing (CBP; Wasserburg et al. 1995) in the parent stars of Z grains, e.g., mixing of envelope material with deep hot regions close to the H-burning shell, resulting in extra productions of $^{13}$C in Z grains. (e.g., Nittler & Alexander 2003). Alternatively, Z grains may come from high-mass stars, where hot bottom burning is activated (Lewis et al. 2013). However, by constraining the efficiency of the reaction $^{22}$Ne($\alpha$,$n$)$^{25}$Mg using the Mo isotopic compositions of Z grains, Liu et al. (2019) showed that current nucleosynthesis models for AGB stars of lower-than-solar metallicity and/or high-masses (> 3 M$_\odot$) cannot explain Z grains' large $^{30}$Si excesses. AB grains (4−5%) have lower $^{12}$C/$^{13}$C ratios (<12) than MS, Y, and X grains (Stephan et al. 2020). They show a similar range of Si isotopic compositions as MS grains, but exhibit a larger range of $^{14}$N/$^{15}$N and higher inferred initial $^{26}$Al/$^{27}$Al ratios. Their isotopic signatures suggest that they may originate in J-type carbon stars (Liu et al. 2017a), born-again AGB stars (Alexander 1993; Amari et al. 2001c), and/or core-collapse supernovae (CCSNe; Liu et al. 2017a). Type X grains have distinct excesses in $^{15}$N, $^{28}$Si, and $^{26}$Al and a wide range of $^{12}$C/$^{13}$C ratios (Fig. 1a). Some of these grains contain evidence for extinct $^{44}$Ti ($t_{1/2}$ = 60 a), which is a proof of their supernova origin (Nittler et al. 1996). Type C grains are also thought to form in CCSNe and are characterized by very large excesses in $^{29}$Si and $^{30}$Si isotopes. In comparison to X grains, C grains are believed to originate from more external CCSN layers (Pignatari et al. 2013b). Putative nova grains (N grains) are generally characterized by large excesses in $^{13}$C, $^{15}$N, $^{30}$Si, and $^{26}$Al, which are likely caused by high-temperature explosive H burning in novae and/or supernovae (e.g., Amari et al. 2001a; Liu et al. 2016; Nittler 2005). Finally, grains with isotopic compositions that do not fit in any of the groups defined above are named ungrouped grains, U grains (<0.1‰), and some of them also likely originated in CCSNe (Liu et al. 2018a; Xu et al. 2015).

The classification of SiC grains has been continuously evolving because of the growing database of presolar grains and the improvements in astrophysical models. For example, it has been suggested to further sub-divide X, AB, and C grains on the basis of various isotopic signatures (Lin et al. 2010; Liu et al. 2016; Liu et al. 2017a). Advances in machine learning algorithms and the availability of the Presolar Grain Database (PGD; (Hynes & Gyngard 2009; Stephan et al. 2020) offer a great opportunity to improve this classification and to gain further insights into the origin of presolar grains. In particular, cluster analysis is a statistical method that enables quantitative determination of groups of samples having similar features based on the density distribution of the dataset. In this study, we investigate the clustering of SiC presolar grains using state-of-the-art cluster analysis techniques and the updated PGD (Stephan et al. 2020). We then compare the results with previous classifications and astronomical models.

## 2. Methods

We used SiC grain data from the PGD, initially compiled ~10 years ago (Hynes & Gyngard 2009) and recently updated (Stephan et al. 2020) (https://presolar.phys-ics.wustl.edu/presolar-grain-database/). The most recent database PGD_SiC_2020-08-18 contains isotopic compositions for 19,759 presolar SiC grains. Here, we worked with two datasets; hereafter DB4 and DB5, which provide the most relevant results for addressing stellar formation environments of presolar grains (see Appendix A). The largest dataset DB4 includes 1354 data points with measured $^{12}$C/$^{13}$C, $^{14}$N/$^{15}$N, $\delta^{29}$Si/$^{28}$Si and $\delta^{30}$Si/$^{28}$Si ratios from PGD_SiC_2020-08-18. DB5 additionally includes inferred initial $^{26}$Al/$^{27}$Al ratios, which reduces the number of observations to 402. For DB5, we considered the database PGD_SiC_2020-01-30 (available throughout almost the entire duration of the project), and additional data published in the last four years (see Appendix A and Supp. Table available on Github1[1]); the combined dataset is similar to the data in the PGD_SiC_2020-08-18, and we, therefore, did not rerun our cluster analysis. We excluded grains of C and U types, since these grains are relatively rare. In addition, since DB4 samples a large number of MS grains, we excluded those having 1σ uncertainties in $\delta^{29}Si/^{28}Si$ and $\delta^{30}Si/^{28}Si$ larger than 10 ‰, to increase the clustering quality. We took the logarithms of all the isotopic ratios, and then scaled them to an average of zero and unit standard deviation. These transformations reduce the skewness of the data and are comparable to the graphical tradition of using equally sized axes, sometimes log-scaled, with optimal minimal and maximal values, to maximize the visualization of data variance. This normalization prevents the clustering from being controlled by the variables having the largest absolute values, rather than the variance of the entire data.

---

[1] https://doi.org/10.5281/zenodo.4304818

Several cluster analysis techniques exist, including clustering approaches based on the distribution or density of the data, connectivity of data points, and average distances between data points (e.g., Jain 2010). Each algorithm presents advantages and disadvantages and can be more or less appropriate to specific datasets. We used a model-based clustering algorithm, which assumes that the dataset is a mixture of probability distributions and finds clusters by maximizing the probability that each data point belongs to a specific cluster. This technique is appropriate for overlapping clusters that are difficult to resolve with other algorithms. The SiC grain groups as currently defined present significant overlap, for example, AB and MS grains have similar Si isotopic compositions, and MS and Z grains have similar ranges of N and C isotopic ratios (Fig. 1a). Therefore, model-based clustering is especially suitable for clustering presolar SiC grains. We used the Mclust R package (Scrucca et al. 2016) that assumes mixtures of Gaussian distributions and selected the best clustering models using the highest Bayesian Information Criterion (BIC) which is known to provide models that best fit overlapping data (Bouveyron et al. 2019). Our R code for presolar SiC cluster analysis is available on GitHub[2].

### 3. Results and Discussion

#### *3.1 Clustering results*

Initial cluster analysis of dataset DB4 yielded, among other clusters, a tightly constrained group of 20 MS grains, having a terrestrial isotopic composition for C, Si, and N. These grains were likely contaminated by Solar System material, and therefore, removed from our datasets, since their presence affects strongly the density distribution of the datasets (Fig. 1b). Table 1, Figs. 2–3, and Table S1 show the properties of the obtained clusters, their isotopic compositions, and linear regression results of their Si isotopic compositions. Both cluster analyses of DB4 and DB5 indicate the presence of nine different clusters of relatively good quality, since all clusters have an average probability close to or higher than 0.8 (Figs. 2d and 3f). However, clusters of grains with overlapping isotopic compositions (in DB4 cluster 1, 3, and 9, hereafter $c1_{DB4}$, $c3_{DB4}$, and $c9_{DB4}$, respectively, and in DB5 clusters $c1_{DB5}$, $c3_{DB5}$, and $c8_{DB5}$, all of which contain mostly MS, Y, and Z grains) have lower ranges of probability than the other clusters. Principal component analysis collapsing the dimensions into the most significant (Fig. 7) shows that using both datasets DB4 and DB5, ~67 % and ~21 % of the dataset variance is included in the first and second components, respectively. While the first component includes the variances of all isotopic ratios, the second is dominated by $^{12}C/^{13}C$, $^{14}N/^{15}N$, and $^{26}Al/^{27}Al$ (Fig. 8). Our results broadly agree with the original classification for the three main groups of presolar SiC grains (AB, X, and MS), but allowed us to discover additional divisions: AB grains are included in two clusters, X in two and three clusters in DB4 and DB5, respectively, and MS in three clusters (see Fig. 8).

---

[2] https://doi.org/10.5281/zenodo.4304837

### 3.2 Mainstream, Y, and Z grains

Mainstream SiC grains are generally accepted to originate from relatively low-mass AGB stars with metallicities comparable to or higher than that of the Sun (Lugaro et al. 2020; Zinner 2014). Their Si isotope ratios mainly reflect the initial compositions of their parent stars, set by Galactic Chemical Evolution (GCE), and possibly slightly modified by dredge-up of $^{30}Si$-rich material that has experienced neutron capture. GCE theory predicts that $^{29,30}Si/^{28}Si$ ratios increase linearly with metallicity (Lewis et al. 2013; Timmes & Clayton 1996), though the details are highly uncertain and local heterogeneities may arise due to inhomogeneous supernova enrichment of star-forming regions (Lugaro et al. 1999; Nittler 2005). The C isotopic ratios of AGB stars depend on their initial composition controlled by GCE and stellar nucleosynthesis, including production of $^{12}C$ by triple alpha reaction during the shell He burning and possible production of $^{13}C$ by CBP and/or hot-bottom burning (HBB) (e.g., Zinner et al. 2006). CBP is a process where envelope material of M<~2.5 $M_\odot$ stars is mixed with deep H-burning shell material (Wasserburg et al. 1995), while HBB happens in massive stars (M>~4 $M_\odot$) for which the bottom of the convective envelope reaches significantly high temperatures to initiate H burning (Nollett et al. 2003). In addition to GCE, the $^{14}N/^{15}N$ ratio is expected to be also set by dredge-up of material that experienced CNO-cycle H burning, although observed ranges in MS, Y, and Z grains are higher than predicted by standard AGB stellar models that do not include CBP and HBB (e.g., Palmerini et al. 2011). Let us consider the MS-rich clusters found here in light of these processes. To aid in the discussion, we follow previous authors (e.g., Amari et al. 2001b; Hoppe et al. 1997; Lewis et al. 2013; Nittler & Alexander 2003) and project the Si-isotope data onto an assumed GCE line, to derive the initial composition ($\delta^{29,30}Si_{init}$) of the parent AGB stars (Figs. 4a–c), and a proxy for the amount of processed material mixed into the envelope during the AGB phase ($\Delta^{30}Si$) (Figs. 4a & c). Here, we considered a GCE line of 1.5 intersecting the solar composition, and a mixing line with a slope of 0.5, following (Nittler & Alexander 2003). Projections on these lines to calculate the initial $\delta^{29}Si$ and the degree of AGB mixing $\Delta^{30}Si$ are shown in Fig. 4a. The specific values for these parameters depend on the precise assumptions made for the projection, but this exercise allows us to investigate the trends qualitatively and see how they vary for the different clusters.

For the DB4 results, cluster c8$_{DB4}$ with 70% Y and 20% Z shows a negative correlation of $\delta^{29}Si_{init}$ and $\Delta^{30}Si$ (see Fig. 4b), in agreement with previous studies (Nittler & Alexander 2003; Zinner et al. 2006), supporting the higher efficiency of third dredge-up and more efficient operation of the reaction $^{22}Ne(\alpha,n)^{25}Mg$ in low metallicity stars because of enhanced maximum stellar temperatures. However, it is noteworthy that this implication is not supported by the Mo isotopic compositions of Z grains (Liu et al. 2019). Our derived divisions rely on a multi-dimensional approach and suggest that the use of $^{12}C/^{13}C = 100$ to separate Y grains from other types of grains is arbitrary. In addition, Si isotopic ratios of the two pure MS clusters $c1_{DB4}$ and $c9_{DB4}$ plot close to the MS line, which translates into a narrow range of $\Delta^{30}Si$ values (Fig. 4c), while cluster $c3_{DB4}$, comprising a significant number of Y grains, has a larger spread in $\Delta^{30}Si$ than the two other clusters. These three clusters show increasing average $\delta^{29,30}Si$ values from $c3_{DB4}$ to $c9_{DB4}$ to $c1_{DB4}$ (Table 1), suggesting increasing ranges of metallicity. Fig. 4c reveals, as previously observed (Hoppe et al. 1996; Huss et al. 1997; Nittler & Alexander 2003), a negative correlation between $\delta^{29}Si_{init}$ and the observed maximum $^{12}C/^{13}C$ for AGB-derived SiC grains, possibly due to increased dredge-up of He-burning material in lower-metallicity stars, or GCE control of initial compositions. Indeed, GCE models predict increasing $^{12}C/^{13}C$ with decreasing metallicity, but the predicted trend seems not to be supported by existing limited astronomical observations (Kobayashi et al. 2020). Fig. 4c also shows the density distribution for MS grains and highlights the very dense region (in yellow) corresponding to the center of cluster $c9_{DB4}$ at $^{12}C/^{13}C \cong 65$. The dense center of $c9_{DB4}$ (narrow ranges of C and Si isotopic compositions) indicates an enhanced production of grains from stars with a narrow range of mass and metallicity. It also shows a second broader high-density region coinciding with the compositional range of $c1_{DB4}$, and the low-density region of heterogeneous cluster $c3_{DB4}$. In addition, available Ti isotopic ratios for these three clusters show smaller ranges for cluster $c9_{DB4}$ compared to $c1_{DB4}$ and $c3_{DB4}$ (e.g., $\delta^{46}Ti/^{48}Ti = 34 \pm 32$ ‰ compared to 59 ±57 ‰ and 13 ± 55 ‰, respectively). Altogether, cluster analysis enables defining four clusters of grains formed in AGB stars with gradually increasing average metallicity: $c8_{DB4}$ (Z- and Y-rich), $c3_{DB4}$ (Y- and MS-rich grains), $c9_{DB4}$ (MS-rich) and $c1_{DB4}$ (MS-rich).

Two puzzles regarding SiC Si isotopes have long been recognized: 1) the slope of the MS SiC Si isotope line is steeper than predicted by GCE models (~1.4 versus 1) and 2) most grains are $^{29,30}Si$-rich, suggesting that their stars are of higher metallicity than the Sun despite forming earlier in Galactic history. A super-solar metallicity origin for large (> 1 μm) MS grains has seen recent support from astronomical observations and trace-element isotope data (Lugaro et al. 2020), but the slope discrepancy remains unexplained, though many models have been suggested to explain it (e.g., Nittler & Dauphas 2006). One possible solution, suggested by Clayton (2003), is that a merger of a dwarf galaxy with the Milky Way a few Ga before Solar System formation could have triggered a burst of star formation and led to a local evolution of Si isotopes down the MS line as low-metallicity gas from the dwarf mixed with the Milky Way disk. Depending on the timing of the merger, stars formed in the starburst of a given mass will have evolved to the AGB phase and produced SiC grains just in time to contribute to the forming Solar System, and an excess of grains from a narrow mass/metallicity range may thus support this model. Heck et al. (2020) recently found that a higher-than-expected fraction of presolar SiC grains have cosmic-ray exposure ages smaller than 300 Ma and argued that this suggests an enhanced star formation around ~7 Ga ago, in agreement with the starburst hypothesis. Alternatively, the Si isotope ratios of MS grains can be simply explained as a result of increasing production of SiC in C-rich AGB stars of higher metallicities, i.e., higher Si abundances in the stellar envelope, as shown by Cristallo et al. (2020). Their calculations, which coupled a chemodynamical GCE model, a dust formation model, and AGB nucleosynthesis models, predict that the Solar System incorporated SiC grains from dying AGB stars in the Solar neighborhood with a restricted range of masses and metallicities. This prediction provides an alternative explanation to the identified compact core in Fig. 4c.

The addition of $^{26}Al/^{27}Al$ data (DB5) yields a significantly smaller sample of MS grains (142 instead of 625 in DB4) for our cluster analysis and hence different clusters for MS grains. This limited sampling may not provide the most accurate clustering results for MS, Y, and Z grains, as extensive data are needed to resolve overlapping compositions. However, interestingly, the analysis reveals two MS-rich clusters, $c1_{DB5}$ and $c8_{DB5}$, with relatively high (low) and low (high) average $^{14}N/^{15}N$ ($^{26}Al/^{27}Al$) ratios, respectively (Fig. 5). This anti-correlation between $^{14}N/^{15}N$ and $^{26}Al/^{27}Al$ ratios in MS grains has not been noted before. AGB models generally predict these ratios should be positively correlated, since H burning at the temperatures experienced by such stars is predicted to produce $^{14}N$ and $^{26}Al$ so that one would expect a positive instead of negative correlation. Higher-temperature H burning in novae and/or some supernovae has been invoked to explain coupled $^{15}N$ and $^{26}Al$ in N and AB grains (Liu et al. 2017a; Liu et al. 2018b), but such burning is not expected for AGB stars. The observed enrichments in $^{26}Al$, $^{15}N$, and scattered Si isotopic ratios of cluster $c8_{DB5}$ may be clues that these grains did not form in AGB stars and/or to future improvements of nucleosynthesis models of AGB stars. However, the two clusters have similar ranges of Si isotopic ratios (see Table 1, Fig. 10), suggesting similar GCE effects.

### 3.3 X, N, and AB grains

We compared N and Al isotopic ratios for our obtained DB5 clusters with recent nova and CCSN stellar nucleosynthesis models in Fig. 5 (Jose & Hernandez 2007; Pignatari et al. 2015). Both N grain-rich cluster $c2_{DB5}$ and the cluster of $^{15}N$-rich AB grains $c6_{DB5}$ match CCSN models by (Pignatari et al. 2015), in which H is ingested into the He/C zone during the supernova explosion, resulting in the production of large amounts of $^{13}C$, $^{15}N$, and $^{26}Al$. In comparison, the other AB cluster $c7_{DB5}$ lies closer in composition to AGB models. These results are in line with those of previous studies, suggesting distinct stellar origins for $^{15}N$-poor AB grains (Liu et al. 2017c) and $^{15}N$-rich grains (Liu et al. 2017c; Liu et al. 2018b). According to astronomical observations, J stars dominantly show $^{14}N/^{15}N$ ratios close-to or greater-than the solar value, in agreement with the range of N isotope ratios of $^{15}N$-poor AB grains (Hedrosa et al. 2013). Hoppe et al. (2019) recently suggested that $^{15}N$-poor AB grains could also have originated from CCSNe that experienced explosive H burning but with a different mixing recipe compared with $^{15}N$-rich AB grains. However, it remains unclear if different mixing scenarios in similar parent supernovae can result in two distinct groups of AB grains as identified by the clustering analysis here. In addition, (Liu et al. 2017a) defined two groups of AB1 ($^{15}N$-rich) and AB2 ($^{15}N$-poor) by using the solar $^{14}N/^{15}N$ ratio as the divider. The N isotopic composition of $c7_{DB5}$ is consistent with this definition of AB2 grains, whereas $c6_{DB5}$ covers a wider range of $^{14}N/^{15}N$ than AB1 grains, up to ~2000, which supports the suggestion of Hoppe et al. (2019) that a fraction of AB2 grains could have originated from CCSNe as AB1 grains. The accurate partitioning using cluster analysis enabled better, defining contours for these two types of grain without having to use the solar isotopic composition as an arbitrary separator. It is interesting to note that our redefined cluster $c6_{DB5}$ yields a better match with the models of Pignatari et al. (2015), specifically for the most $^{26}$Al-depleted compositions. However, note that the CCSN mixtures with $^{14}$N/$^{15}$N ratios above the solar value are O-rich (Liu et al. 2017a), and equilibrium condensation calculations do not predict SiC condensation under such conditions. Thus, if these grains are indeed from CCSNe, they may have formed by a non-equilibrium process (e.g. Deneault 2017).

X grains are clustered in two and three clusters using DB4 and DB5, respectively. Cluster $c5_{DB4}$ or $c6_{DB5}$ have strongly correlated Si isotopes similar to grains from previously proposed subtype X1 (Lin et al. 2002). The other X grains, defined as X0 and X2 by Lin et al. (2010), are either all clustered in the heterogeneous $c4_{DB4}$, or split between a heterogeneous cluster $c4_{DB5}$ and cluster $c10_{DB5}$ with a large range of $^{12}$C/$^{13}$C and stronger depletions in $^{29}$Si and $^{30}$Si abundances, respectively. However, cluster $c10_{DB5}$ does not have a distinct range in $^{26}$Al/$^{27}$Al and overlaps with $c5_{DB5}$ (Fig. 5). Moreover, the data distribution for all X grains is the same in DB4 and DB5 (Fig. 6), and the use of the larger dataset (DB4) did not lead to their splitting. For these reasons, the additional cluster $c10_{DB5}$ may not be significant. Based on previous CCSN models the tightly correlated Si isotopic ratios of cluster c5 X grains suggest mixing of material from the inner S/Si (Hoppe et al. 2010; Rauscher et al. 2002) (or Si/C zone in the models of Pignatari et al., 2013a) and outer He/C zones. The scattered cluster c4 likely indicates additional contributions from other shells such as the O-rich shells having Si isotopic compositions strongly deviating from the trend defined by cluster c5 X grains in the Si 3-isotope plot, as previously suggested by Hoppe et al. (2010).

## 4. Conclusions

Our analysis shows that with the available data on presolar SiC grains, clustering based on C, N, Si isotopic compositions (dataset named DB4), and additional Al isotopic ratios (DB5) enables us to accurately define divisions between different groups of grains. Our conclusions are summarized below.

- Four clusters of grains formed in AGB stars with gradually increasing metallicity: $c8_{DB4}$ (with mainly Z and Y grains), $c3_{DB4}$ (predominantly Y and MS grains), $c9_{DB4}$ (almost all MS grains), and $c1_{DB4}$ (MS grains).

- Cluster $c9_{DB4}$, containing 1/3 of all MS grains, has a very narrow range of $^{12}$C/$^{13}$C and initial $\delta^{29,30}$Si ratios, indicating that these grains came from parent AGB stars with a narrow range of masses and metallicities. This inference is consistent with an enhanced grain production from a starburst event prior to the Solar System formation but is more likely to be a natural consequence of GCE and preferred SiC production in high-metallicity C-rich AGB stars.

- Adding inferred initial $^{26}$A/$^{27}$Al data to the clustering identifies a cluster of MS grains with enrichments in $^{26}$Al and $^{15}$N, which questions their supposed AGB stellar origins or points to problems in current AGB models. However, since the MS data on their Al isotopic ratios remains limited, we stress that future studies should give special attention to $^{26}Al/^{27}Al$ when analyzing MS grains.

- Two AB-rich clusters have either low $^{14}N/^{15}N$ and high $^{26}Al/^{27}Al$, or high $^{14}N/^{15}N$ and low $^{26}Al/^{27}Al$. Their comparable ranges of Si isotopic compositions suggest that they originated from parent stars with similar metallicities, and thus a GCE-related explanation for their difference N and Al isotopic compositions is unlikely. Comparing their compositions with nova and CCSN models points to an origin in CCSNe for $^{15}N$ and $^{26}Al$-rich AB grains, specifically with mixtures from outer H envelope material and explosive H-burning products in the inner He/C zone.

- A cluster mainly made of N grains and two clusters of X grains, one of which exhibits very correlated Si isotopic ratios, suggests mixing of material from the inner S/Si (or Si/C) and outer He/C zones. One of the two X grain clusters shows a heterogeneous Si composition and deviates from the 2/3 line in the Si 3-isotope plot, which indicates mixing with material from the other O-rich zones, as previously suggested.

Our results demonstrate the power of cluster analysis to separate varying stellar formational environments of presolar SiC grains using the complex covariance of their isotopic compositions. It also highlights the need to expand the PGD and clustering to include other attributes, e.g., isotope ratios of additional elements, morphological features, interstellar ages. These additions would likely provide new insights into stellar nucleosynthesis, GCE and interstellar processes and improve the classification of presolar grains.

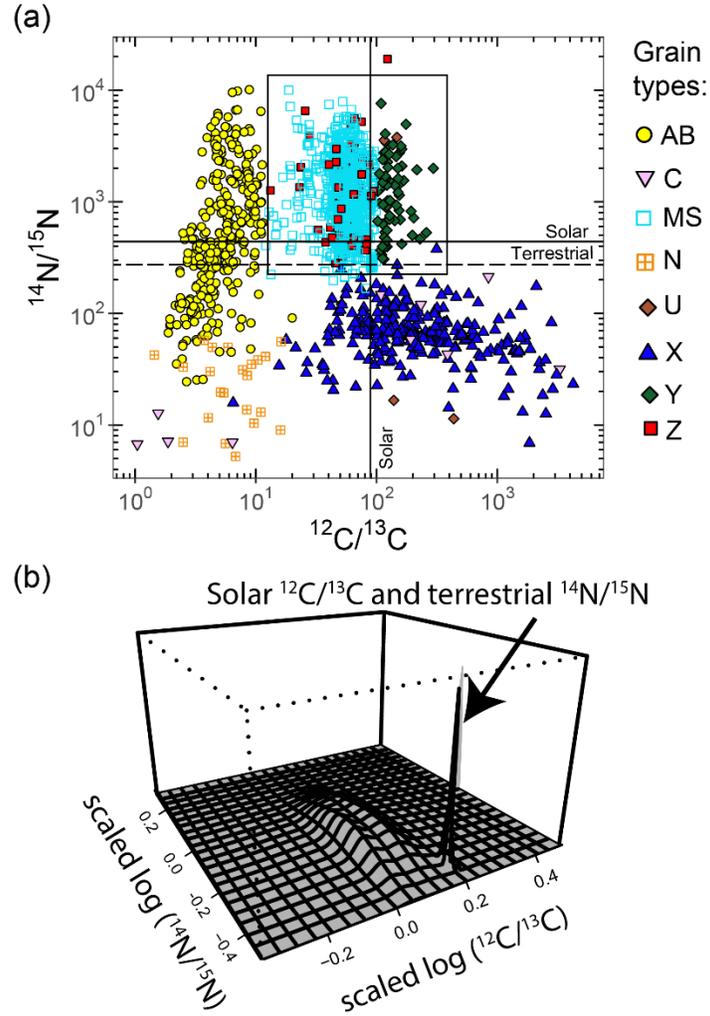

**Figure 1.** Nitrogen and carbon (a) isotopic compositions of presolar SiC grains. Plotted data are from the newly updated PGD (Stephan et al. 2020). We used the following standard isotopic ratios: $(^{12}C/^{13}C)_{solar} = 89$ (Lambert & Mallia 1968), $(^{14}N/^{15}N)_{solar} = 440$ (Marty et al. 2011), $(^{14}N/^{15}N)_{terrestrial} = 272$ (Junk & Svec 1958). (b) is a perspective plot of the bivariate density distribution, scaled to an average of zero and unit standard deviation, for the area shown with the square in (a). Grains likely affected by Solar System contamination create a peak at the solar $^{12}C/^{13}C$ and terrestrial $^{14}N/^{15}N$ isotopic ratios (b).

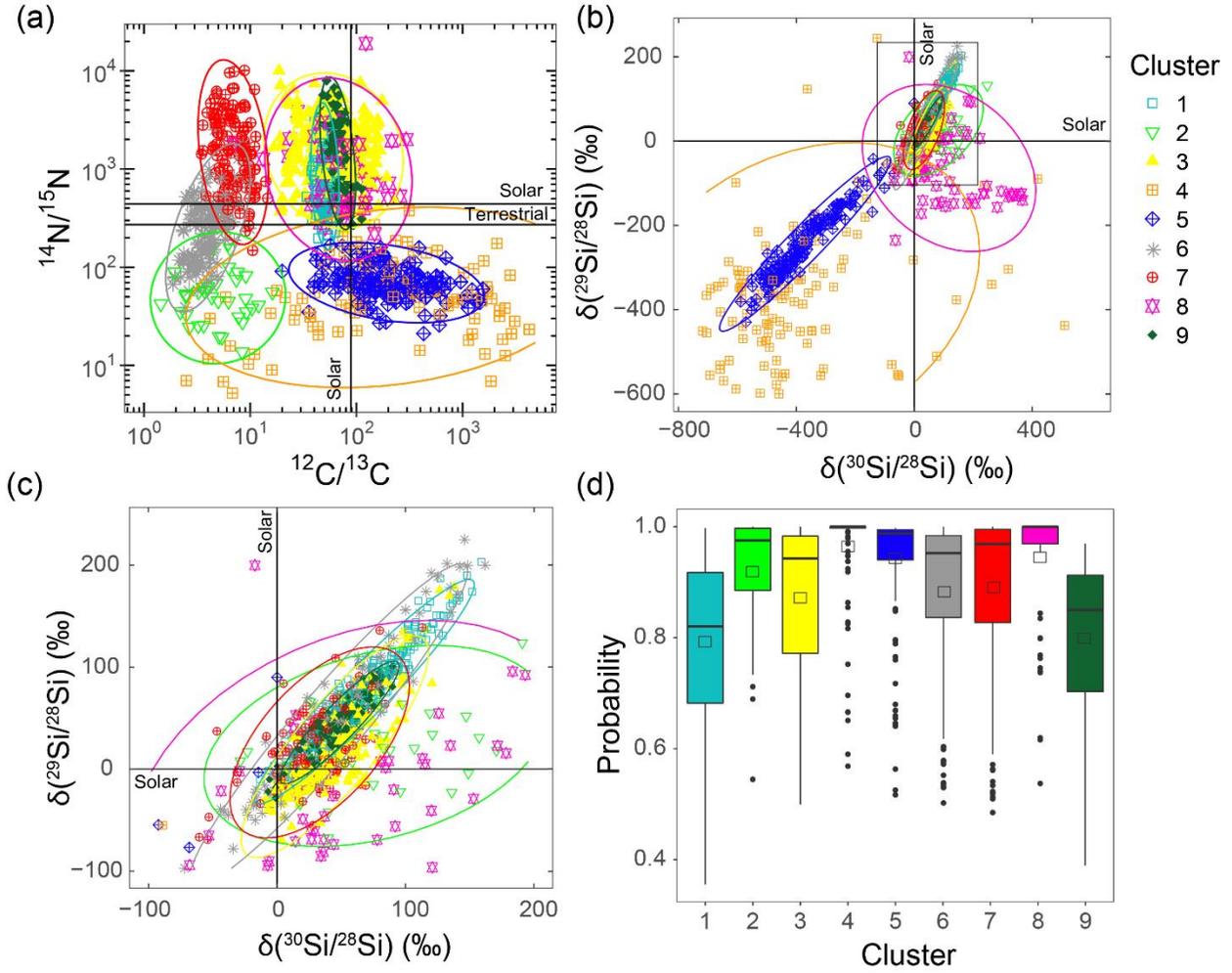

**Figure 2.** (a) Nitrogen, carbon (a) and silicon (b and c) isotopic compositions of clustered presolar SiC grains for DB4 (subset of the PGD (Stephan et al. 2020)), where (c) is a zoom-in of (b). Ellipses are guidelines showing the different clusters. (d) Boxplot showing the probability that data points belong to their clusters. Open squares and horizontal lines are average and median probabilities, respectively, and black dots are outliers.

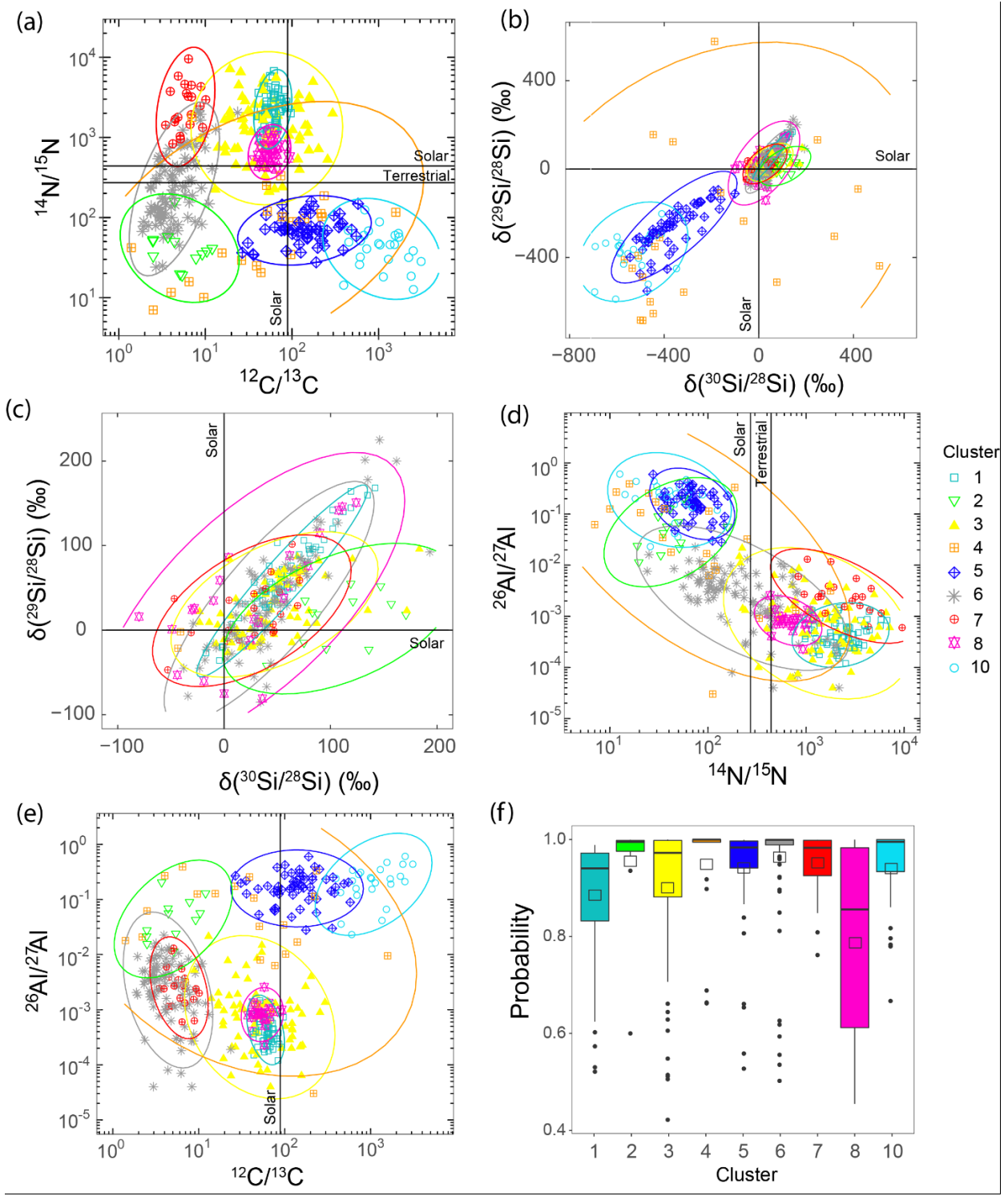

**Figure 3.** Same as Fig. 2. Here cluster analysis is conducted with DB5. (d) and (e) show their inferred initial $^{26}$Al/$^{27}$Al as a function of their $^{14}$N/$^{15}$N and $^{12}$C/$^{13}$C ratios, respectively.

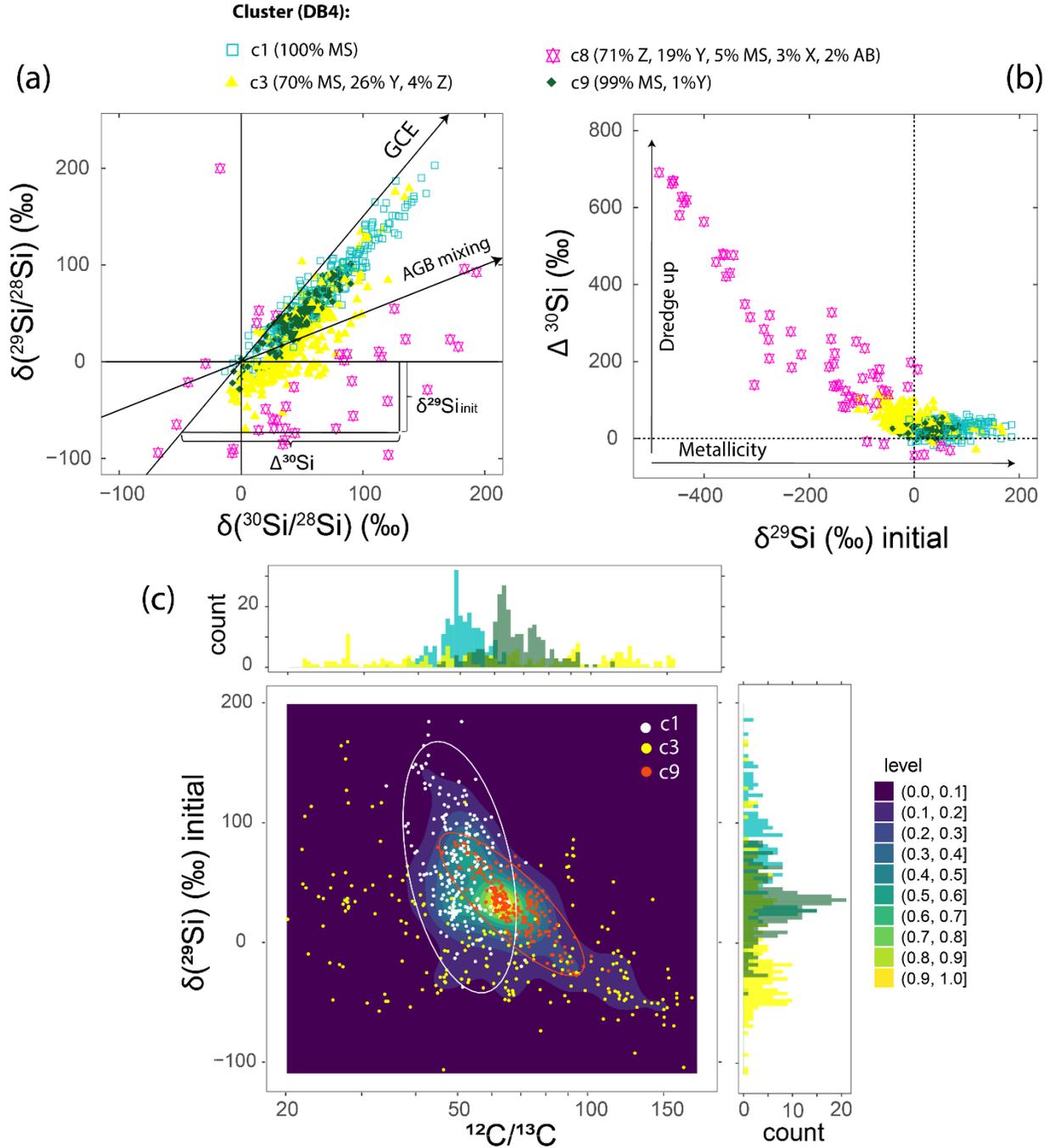

**Figure 4.** (a) Si isotopic ratios for SiC grains from clusters $c1_{DB4}$, $c3_{DB4}$, $c8_{DB4}$, and $c9_{DB4}$, which comprise MS, Y and Z grains and are believed to have formed in AGB stars. Solid arrows show the GCE trend we assumed to represent the initial composition of the parent star, and the AGB mixing line, reflecting changes in grain composition due to dredge-up events (Nittler & Alexander 2003). Here, we considered a GCE line of 1.5 and intersecting the solar composition, and a mixing line with a slope of 0.5, following (Nittler & Alexander 2003). The initial $\delta^{29}Si$ and the degree of AGB mixing (estimated with $\Delta^{30}Si$) were calculated using these lines and are shown in (b) and (c) with C isotopic ratios. In (c), we also plotted the density distribution of the data

(highest and lowest in yellow and dark blue areas), and histograms for $^{12}C/^{13}C$ and initial $\delta^{29}Si$ for the three selected clusters. White and orange ellipses are guidelines showing clusters $c1_{DB4}$ and $c9_{DB4}$, respectively.

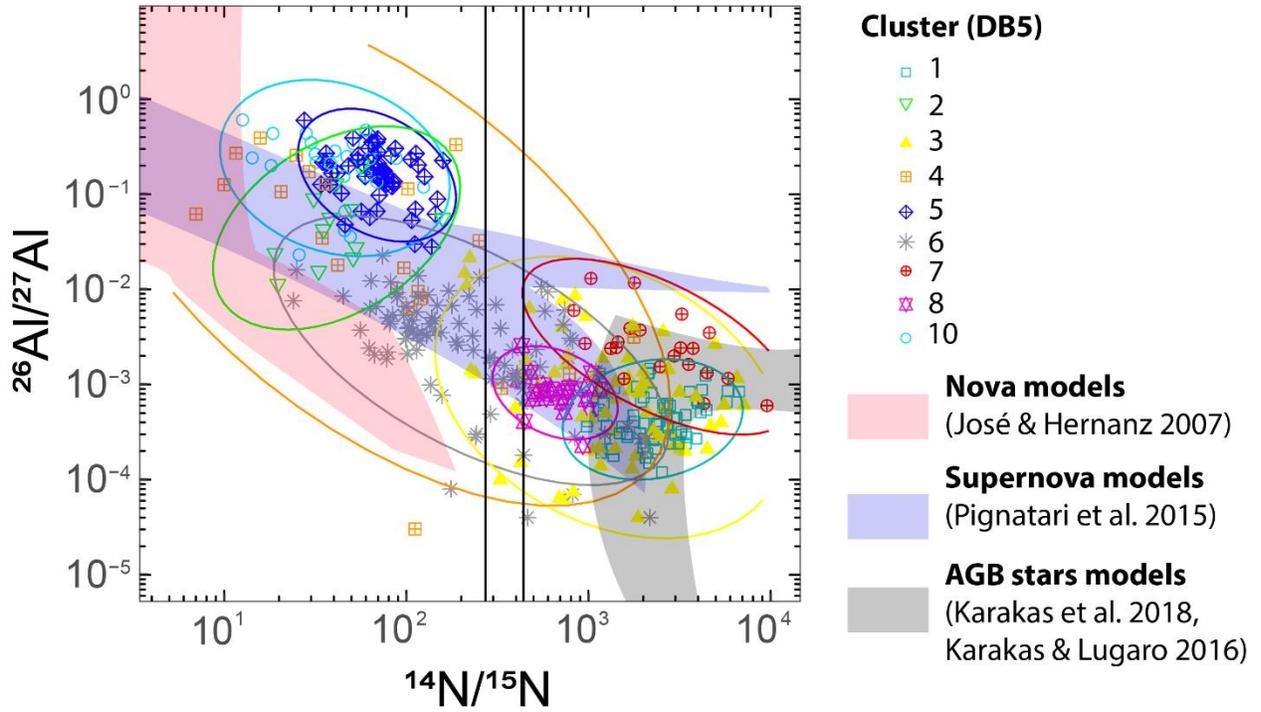

**Figure 5.** Al and N isotopic ratios for SiC grains clusters using DB5 compared to compositions of grains from nova and supernova nucleosynthesis models (Jose et al. 2004; Karakas & Lugaro 2016; Karakas et al. 2018; Pignatari et al. 2015). Details on modeled stars in (d) are given in Karakas & Lugaro (2016), Karakas et al. (2018), and Liu et al. (2017b).

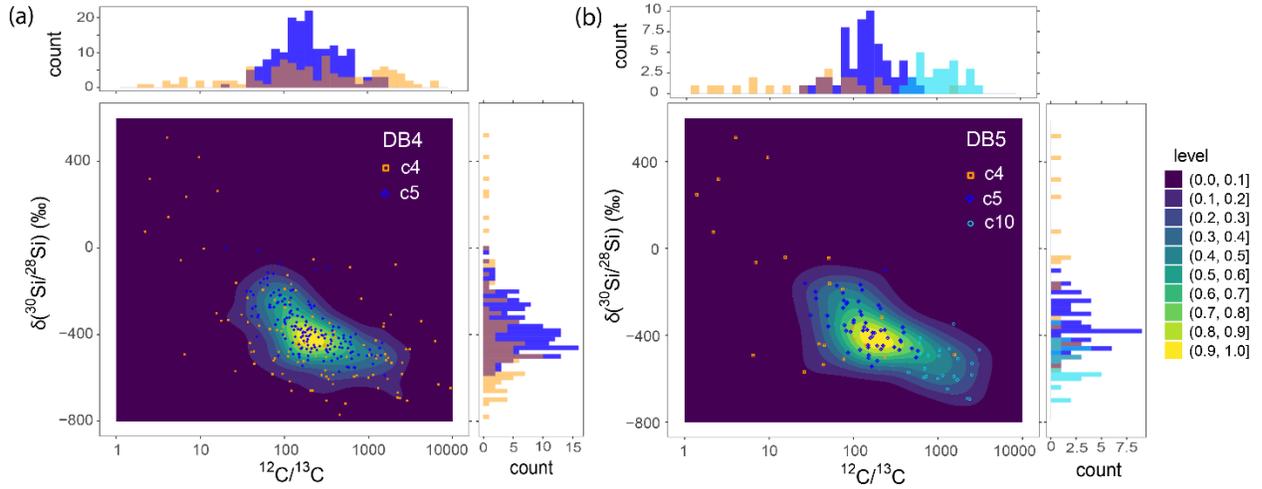

**Figure 6.** $\delta^{30}Si/^{28}Si$ and $^{12}C/^{13}C$ for X grain-bearing clusters from DB4 (a) and DB5, and their density distribution (highest density shown in yellow) (b). Histograms for these two variables are also shown. $\delta^{30}Si/^{28}Si$ and $^{12}C/^{13}C$ are correlated as previously noticed by (Lin, et al. 2010). Since DB4 has more data than DB5 and both datasets show the same distribution, the splitting between $c5_{DB5}$ and $c10_{DB5}$ may not be significant.

**Table 1** Isotopic data of clusters found with DB4 and DB5.

| Cluster | N | Types | $^{12}C/^{13}C$ | | $^{14}N/^{15}N$ | | $\delta^{29}Si/^{28}Si$ (‰) | | $\delta^{30}Si/^{28}Si$ (‰) | | $^{26}Al/^{27}Al$ | | Si isotope correlation[a] | | | | | | |
|---|---|---|---|---|---|---|---|---|---|---|---|---|---|---|---|---|---|---|---|
| DB4 | | | Mean | Std Dev | Mean | Std Dev | Mean | Std Dev | Mean | Std Dev | Mean | Std Dev | R² | F | Intercept | Std error | Slope | Std error | p value |
| 1[b] | 229 | 100% MS | 51 | 6 | 1080 | 804 | 76 | 45 | 68 | 35 | | | 0.91 | 2271 | -7.2 | 2 | 1.22 | 0.03 | < 2e-16 |
| 2 | 34 | 62% AB 38% N | 6 | 3 | 59 | 39 | 26 | 42 | 82 | 59 | | | 0.23 | 9 | -1.9 | 11 | 0.34 | 0.11 | 4E-03 |
| 3 | 220 | 70% MS 26% Y 4% Z | 73 | 42 | 1995 | 1518 | 22 | 44 | 45 | 30 | | | 0.71 | 528 | -34.5 | 2.9 | 1.26 | 0.06 | < 2e-16 |
| 4 | 110 | 91% X 9% N | 788 | 1424 | 67 | 59 | -394 | 163 | -389 | 273 | | | 0.1 | 12 | -321 | 26 | 0.19 | 0.05 | 8.E-04 |
| 5 | 166 | 99.4% X 0.6% AB | 283 | 288 | 75 | 29 | -244 | 84 | -371 | 118 | | | 0.92 | 1975 | 11 | 6 | 0.69 | 0.02 | < 2e-16 |
| 6 | 154 | 100% AB | 4 | 2 | 351 | 346 | 43 | 64 | 40 | 44 | | | 0.91 | 1456 | -12.8 | 2.2 | 1.39 | 0.04 | < 2e-16 |
| 7 | 139 | 98.6% AB 1.4% MS | 7 | 2 | 2133 | 1912 | 26 | 38 | 33 | 28 | | | 0.35 | 75 | 0.21 | 4 | 0.8 | 0.09 | 1E-14 |
| 8 | 63 | 71% Z 19% Y 5% MS 3% X 2% AB | 82 | 55 | 1542 | 2439 | -64 | 79 | 118 | 118 | | | 0.07 | 5 | -12.7 | 13.8 | -0.18 | 0.08 | 4E-02 |
| 9 | 239 | 99% MS 1% Y | 68 | 11 | 1859 | 1331 | 46 | 24 | 49 | 19 | | | 0.84 | 1247 | -12.2 | 1.8 | 1.19 | 0.03 | < 2e-16 |
| DB5 | | | | | | | | | | | | | | | | | | | |
| 1[b] | 70 | 100% MS | 61 | 13 | 2496 | 1194 | 58 | 45 | 55 | 32 | 0.0005 | 0.0003 | 0.9 | 637 | -15 | 3 | 1.34 | 0.05 | < 2e-16 |
| 2 | 12 | 75% N 25% AB | 6 | 3 | 49 | 37 | 14 | 32 | 109 | 39 | 0.06 | 0.06 | 0.23 | 3 | -28.6 | 25.9 | 0.39 | 0.22 | 0.12 |
| 3 | 69 | 67% MS 28% Y 3% AB 3% Z | 71 | 60 | 1839 | 1553 | 30 | 34 | 47 | 39 | 0.002 | 0.004 | 0.24 | 21 | 9.8 | 5.8 | 0.43 | 0.09 | 9e-10 |
| 4 | 23 | 61% X 22% N 9% AB 9% MS | 133 | 324 | 187 | 385 | -274 | 323 | -214 | 333 | 0.09 | 0.12 | 0.09 | 2 | -211 | 79 | 0.3 | 0.2 | 0.16 |
| 5 | 56 | 100% X | 176 | 126 | 75 | 30 | -263 | 100 | -358 | 106 | 0.19 | 0.11 | 0.62 | 87 | 4.6 | 29.8 | 0.75 | 0.08 | 8e-13 |
| 6 | 100 | 98% AB 2% Z | 5 | 3 | 383 | 459 | 33 | 57 | 42 | 40 | 0.004 | 0.004 | 0.67 | 199 | -16.5 | 4.9 | 1.18 | 0.08 | < 2e-16 |
| 7 | 21 | 100% AB | 6 | 1.8 | 2950 | 2068 | 23 | 34 | 26 | 35 | 0.003 | 0.003 | 0.28 | 8 | 9.6 | 8 | 0.5 | 0.2 | 0.013 |
| 8 | 26 | 92% MS 8% Z | 56 | 13 | 686 | 213 | 27 | 73 | 22 | 58 | 0.0009 | 0.0004 | 0.32 | 11 | 11.7 | 13 | 0.71 | 0.21 | 3e-3 |
| 10 | 25 | 100% X | 1283 | 763 | 47 | 25 | -374 | 86 | -531 | 89 | 0.24 | 0.14 | 0.09 | 2 | -216 | 104 | 0.3 | 0.19 | 0.14 |

[a] Statistical results of the linear regression of their $\delta^{29,30}Si$ isotopic ratios.
[b] Clusters in DB4 and DB5 were named similarly based on the co-occurrence of similar grains in each of the two considered clusters. Clusters c9$_{DB4}$ and c10$_{DB5}$ are named differently because they do not have corresponding clusters in DB5 and DB4, respectively.

**Acknowledgments:**

We thank Maria Lugaro and Maximilien Verdier-Paoletti for fruitful discussions. Data-driven studies of mineral evolution and mineral ecology have been supported by the Alfred P. Sloan Foundation, the W. M. Keck Foundation, the John Templeton Foundation (grant #60645), the NASA Astrobiology Institute ENIGMA team (80NSSC18M0093), a private foundation, and the Carnegie Institution for Science. T.S. was supported by NASA through grants 80NSSC17K0250 and 80NSSC17K0251. N.L. acknowledges financial support from NASA 80NSSC20K0387. Any opinions, findings, or recommendations expressed herein are those of the authors and do not necessarily reflect the views of the National Aeronautics and Space Administration.

**Appendix A: Additional details on methods**

Clustering with DB5 was performed with the former database PGD_SiC_2020-01-30 (available before the lately updated one and throughout almost the entire duration of the project), and additional data published in the last four years (Hoppe et al. 2018; Hoppe et al. 2019; Liu et al. 2018a; Liu et al. 2016; Liu et al. 2017a; Liu et al. 2017b; Liu et al. 2017c; Liu et al. 2018b; Liu et al. 2019; Nguyen et al. 2018). The very recently updated database PGD_SiC_2020-08-18 contains only one additional initial $^{26}Al/^{27}Al$ ratio, so it is very similar to the one considered here. The supplementary Table available on Github[3] gives the datasets DB4 and DB5 used in this study and probabilities that each data point belongs to clusters from DB4 and DB5.

Although clustering with three attributes ($^{12}C/^{13}C$, $^{29}Si/^{28}Si$, and $^{30}Si/^{28}Si$) would increase the number of grains to 18172, this cluster analysis yields unstable results with clusters varying significantly at each run. The addition of the Ti isotopic ratios produces a dataset of 382 grains. Its cluster analysis yields 3 clusters compared to 6 grain types (M, AB, X, Y, Z, N) from the original classification, suggesting the need for more measurements of isotopic ratios of presolar SiC grains, including Ti isotopes. Therefore, in this study, we provide results for the two datasets DB4 and DB5, which yield the most relevant results for addressing formation environments of presolar SiC grains. It should be noted that each dataset has different proportions of grain types, which may lead to sample biases. While this problem is beyond the scope of the present study, future work should address the effect of these sampling biases on cluster analysis.

**Appendix B: Principal component analysis.**

We performed a principal component analysis (PCA), allowing to better visualize the variance of the data in a dimensionally reduced space. PCA biplots for DB4 and DB5 are shown in Figure 7. Figure 8 shows the contribution of each variable used in cluster analysis in the first and second principal components, for DB4 (a and b) and DB5 (c and d).

---

[3] https://doi.org/10.5281/zenodo.4304818

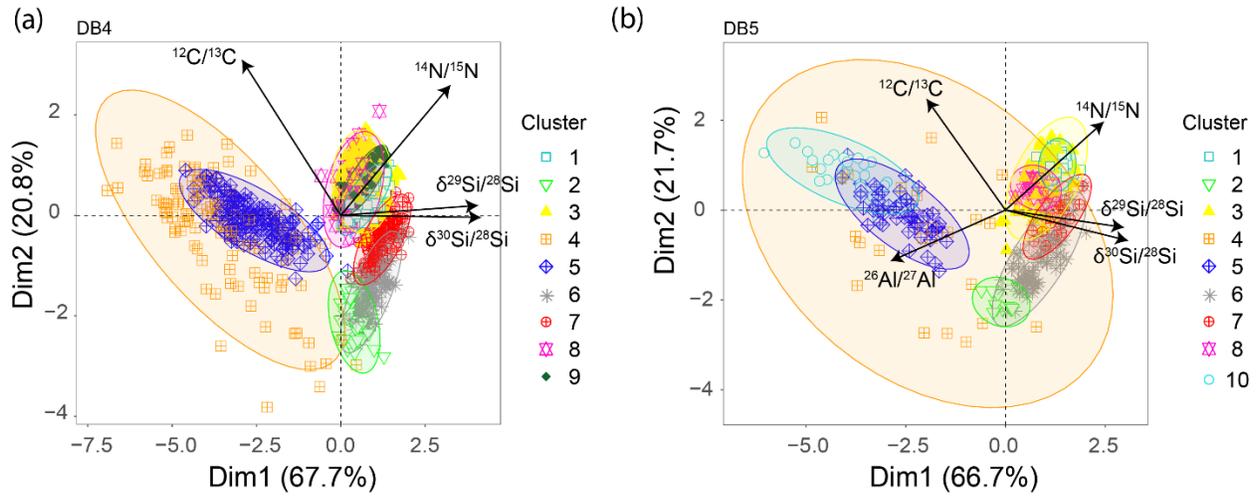

**Figure 7.** Biplot showing the contribution of the four considered features $^{12}C/^{13}C$, $^{14}N/^{15}N$, $^{29}Si/^{28}Si$, and $^{30}Si/^{28}Si$ in DB4 (a) and additional $^{26}Al/^{27}Al$ (b) in the two first principal components and the distribution of data in the collapsed dimensional space.

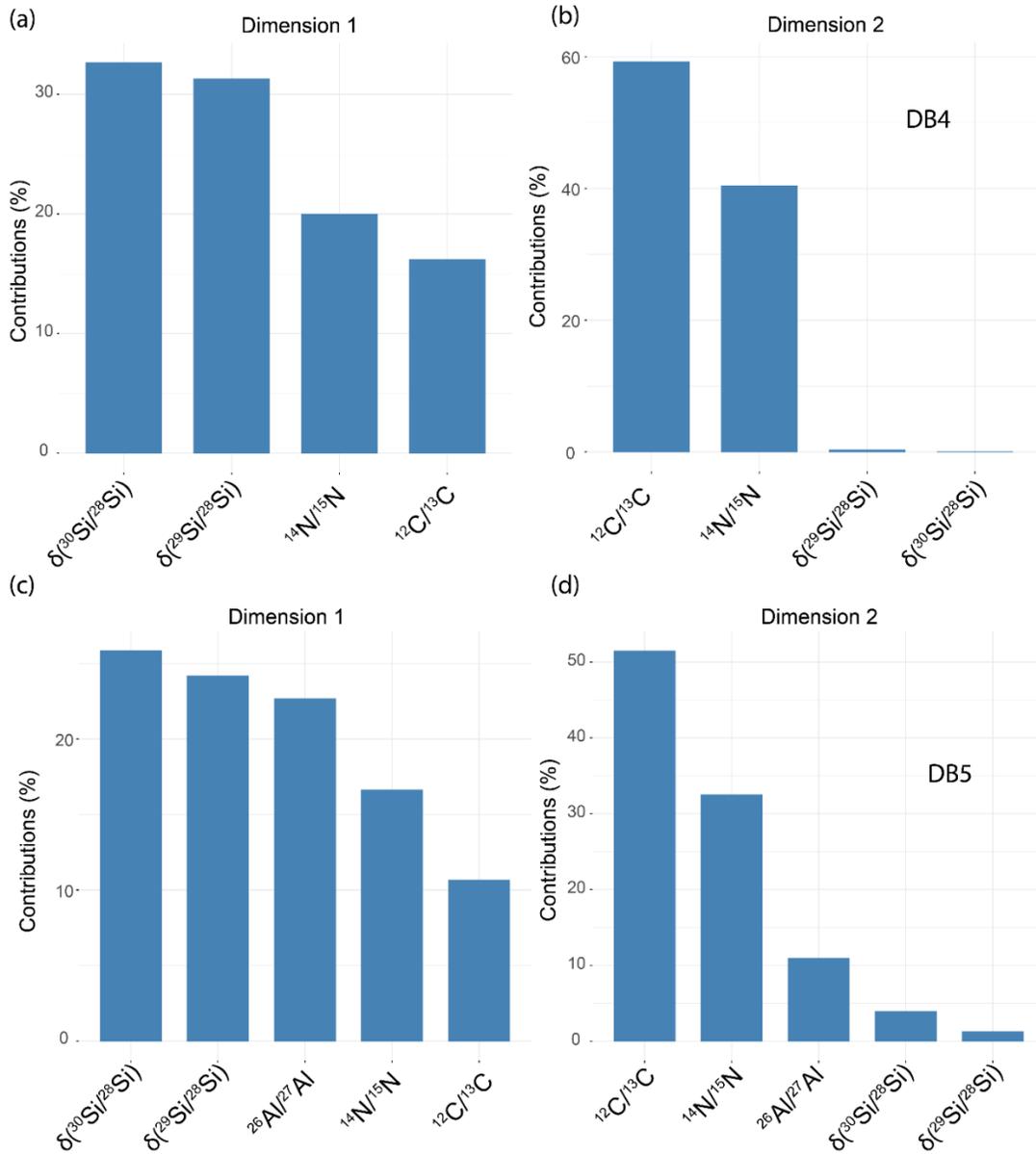

**Figure 8.** Contribution of isotopic ratios to the first and second principal components ((a) and (b), respectively) for DB4, and for DB5 ((c) and (d), respectively).

## Appendix C: Comparison of clusters and grain types for DB4 and DB5

Figure 9 shows confusion matrices as bar plots, comparing clusters and grain types. Average compositions of clusters from DB4 and DB5 are compared in Fig. 10.

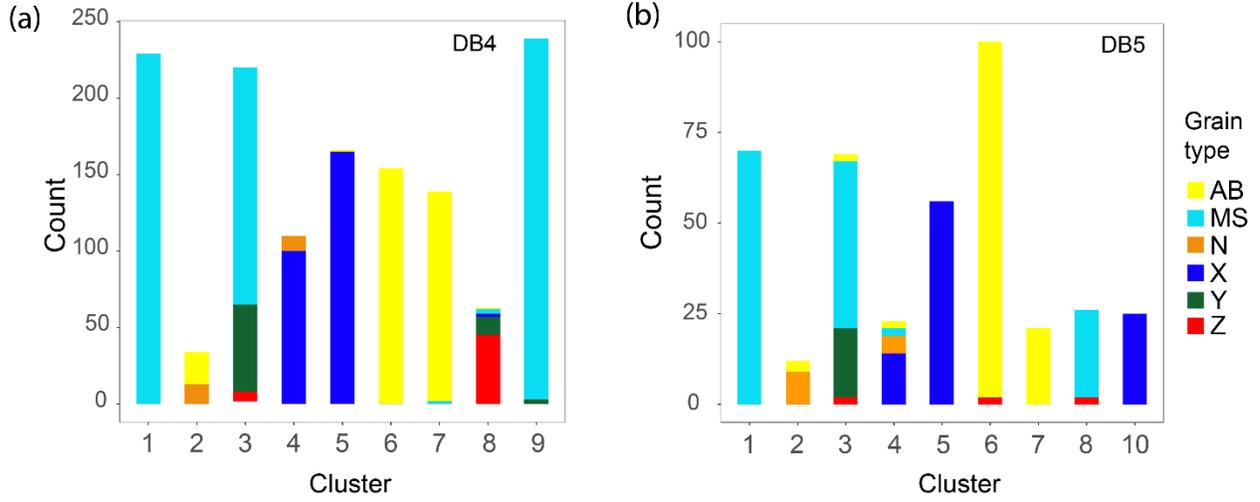

**Figure 9.** Bar plots comparing clusters and grain types for DB4 (a) and DB5 (b).

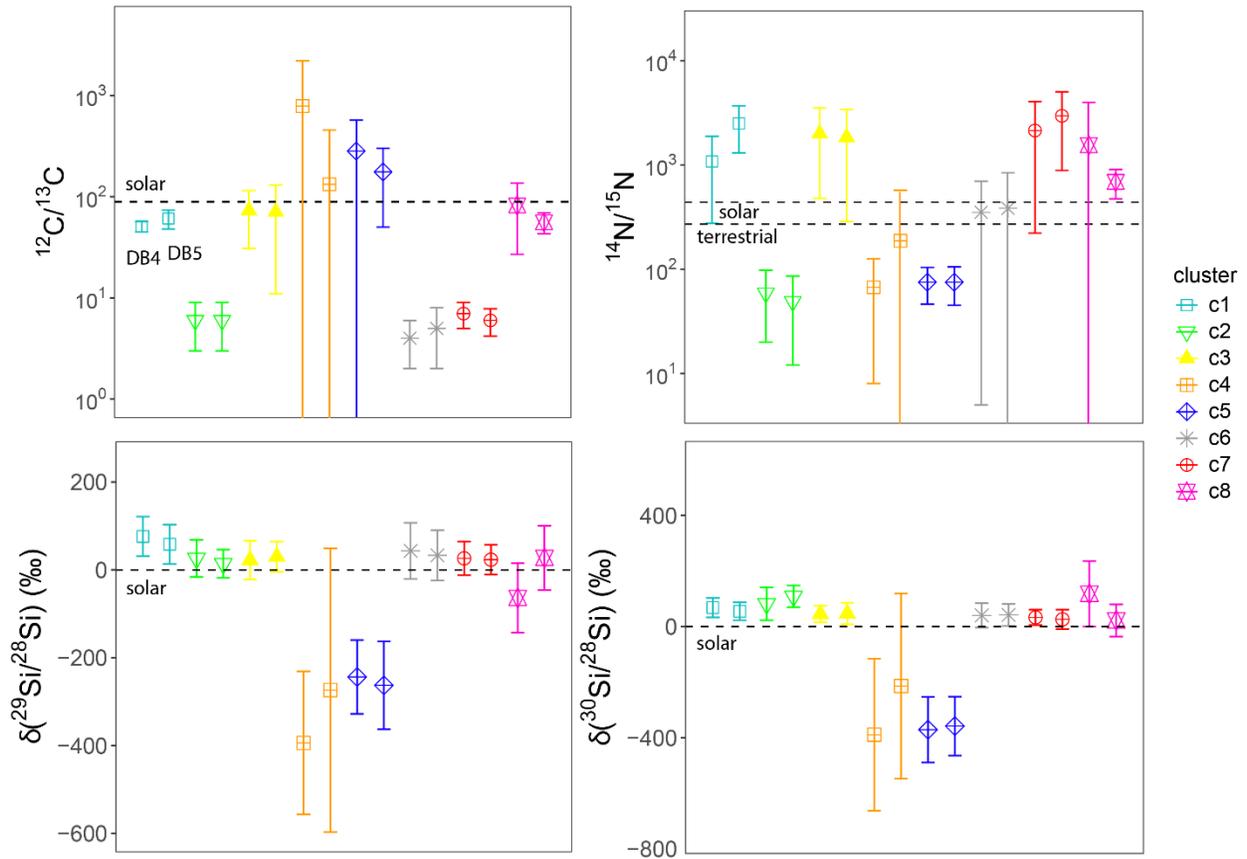

**Figure 10.** Comparison between average isotopic compositions of clusters from DB4 (left symbol) and DB5 (right symbol, as shown for cluster c1 in the $^{12}$C/$^{13}$C plot). Error bars are 1σ standard deviations for each cluster.